\documentclass[12pt]{article}
\usepackage{graphicx}

\def\pbnr{}
\def\speaker{Vasily Ryadovikov}
\def\onbehalfof{SVD-2 Collaboration}
\def\title{Detection of $D^{\pm}$ mesons production\\ in pA-interactions at 70 GeV}
\def\affiliation{IHEP, Protvino, Moscow region, Russia\\E-mail: riadovikov@ihep.ru}
\def\support{The workshop was supported by the University of Manchester, IPPP, STFC, 
and IOP}

\newcommand\pubnumber{\pbnr}
\newcommand\pubdate{\today}
\def\Title#1{\begin{center} {\Large #1 } \end{center}}
\def\Author#1{\begin{center}{ \sc #1} \end{center}}

\newcommand{\OnBehalf}[1]{\sbox0{#1}\ifdim\wd0=0pt
        {}
	\else
	{\\on behalf of #1}
	\fi}
\newcommand{\SupportedBy}[1]{\sbox0{#1}\ifdim\wd0=0pt
        {}
	\else
	{\footnote{#1}}
	\fi}
\def\Address#1{\begin{center}{ \it #1} \end{center}}

\newcommand\pubblock{\includegraphics[width=5cm]{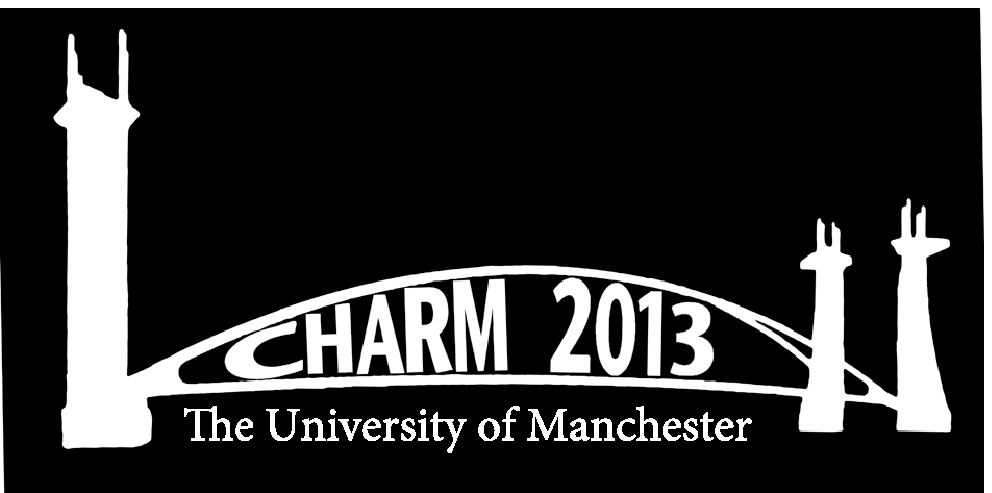}\hfill{\begin{tabular}{l} \pubnumber\\
         \pubdate  \end{tabular}}}
\newenvironment{Abstract}{\begin{quotation}  }{\end{quotation}}
\newenvironment{Presented}{\begin{quotation} \begin{center} 
             PRESENTED AT\end{center}\bigskip 
      \begin{center}\begin{large}}{\end{large}\end{center} \end{quotation}}

\def\venue{The 6$^{th}$ International Workshop on Charm Physics\\
(CHARM 2013)\\
Manchester, UK,  31 August -- 4 September, 2013}

\def\beq{\begin{equation}}
\def\eeq#1{\label{#1}\end{equation}}
\def\eeqn{\end{equation}}

\def\beqa{\begin{eqnarray}}
\def\eeqa#1{\label{#1}\end{eqnarray}}
\def\eeqan{\end{eqnarray}}

\let\bar=\overbar

\def\Dslash{\not{\hbox{\kern-4pt $D$}}}
\def\dslash{\not{\hbox{\kern-2pt $\del$}}}

\def\msb{{\bar{\ssstyle M \kern -1pt S}}}

\begin{document}
\begin{titlepage}
\pubblock

\vfill
\Title{\title}
\vfill
\Author{\speaker\SupportedBy{\support}\OnBehalf{\onbehalfof}}
\Address{\affiliation}
\vfill
\begin{Abstract}
The results of analysis SERP-E-184 experiment \cite{E184} data, obtained with 70 GeV 
proton beam irradiation of active target with carbon, silicon and lead plates 
are presented. For 3-prongs charged charmed mesons decays, event selection 
criteria were developed and detection efficiency was calculated with detailed 
simulation using FRITIOF7.02 and GEANT3.21 programs. Signals of decays were found 
and charm production inclusive cross sections estimated at near threshold 
energy.  The lifetimes and A-dependence of cross section were measured. Yields 
of D mesons and their ratios in comparison with data of other experiments and 
theoretical predictions are presented.
\end{Abstract}
\vfill
\begin{Presented}
\venue
\end{Presented}
\vfill
\end{titlepage}
\def\thefootnote{\fnsymbol{footnote}}
\setcounter{footnote}{0}
%

\section{Introduction}
The open charm production cross section $\sigma(c\bar c)$ at near threshold energy 
in pA-interactions was given in our earlier works \cite{PAN73, PAN74} on 
research of 
$D^0$ mesons characteristics with SVD-2 setup \cite{IET56}. 
In this work the results 
of search and analysis of $D^+ \rightarrow K^-\pi^+\pi^+$ 
and $D^- \rightarrow K^+\pi^-\pi^-$ decays 
in pA-interactions at 70 GeV are presented. Charged charm mesons production 
inclusive cross sections were estimated and their properties were measured.
52 million of inelastic events were detected with three nuclear targets and 
used in our analysis. Selection procedures for events with possible 3-prongs 
decays of charged $D$ mesons were the following:\\
- Reconstruction of tracks and primary vertex using vertex detector data.\\ 
- Search of two prongs secondary vertices in track parameters space \verb|{a, b}| 
\cite{Pr17}.  
In this space each track is presented by a point and all points for tracks 
from the same vertex lie on a straight line.\\ 
- Spatial reconstruction of charged particles tracks in the magnetic 
spectrometer.\\
- Search of 3-prongs secondary vertices taking into account their charge signs 
and spatial association with primary vertex.

After primary selection there were 16320 events with ($K^-\pi^+\pi^+$) decay hypothesis 
and 8439 with ($K^+\pi^-\pi^-$) hypothesis. In fig. 1 raw experimental effective 
mass spectra of two systems are presented. Signals from $D^\pm$ mesons 
can be observed over the large background. To diminish this background additional cuts 
and modeling were required. 
\begin{figure*}[h]
\includegraphics[scale=0.35]{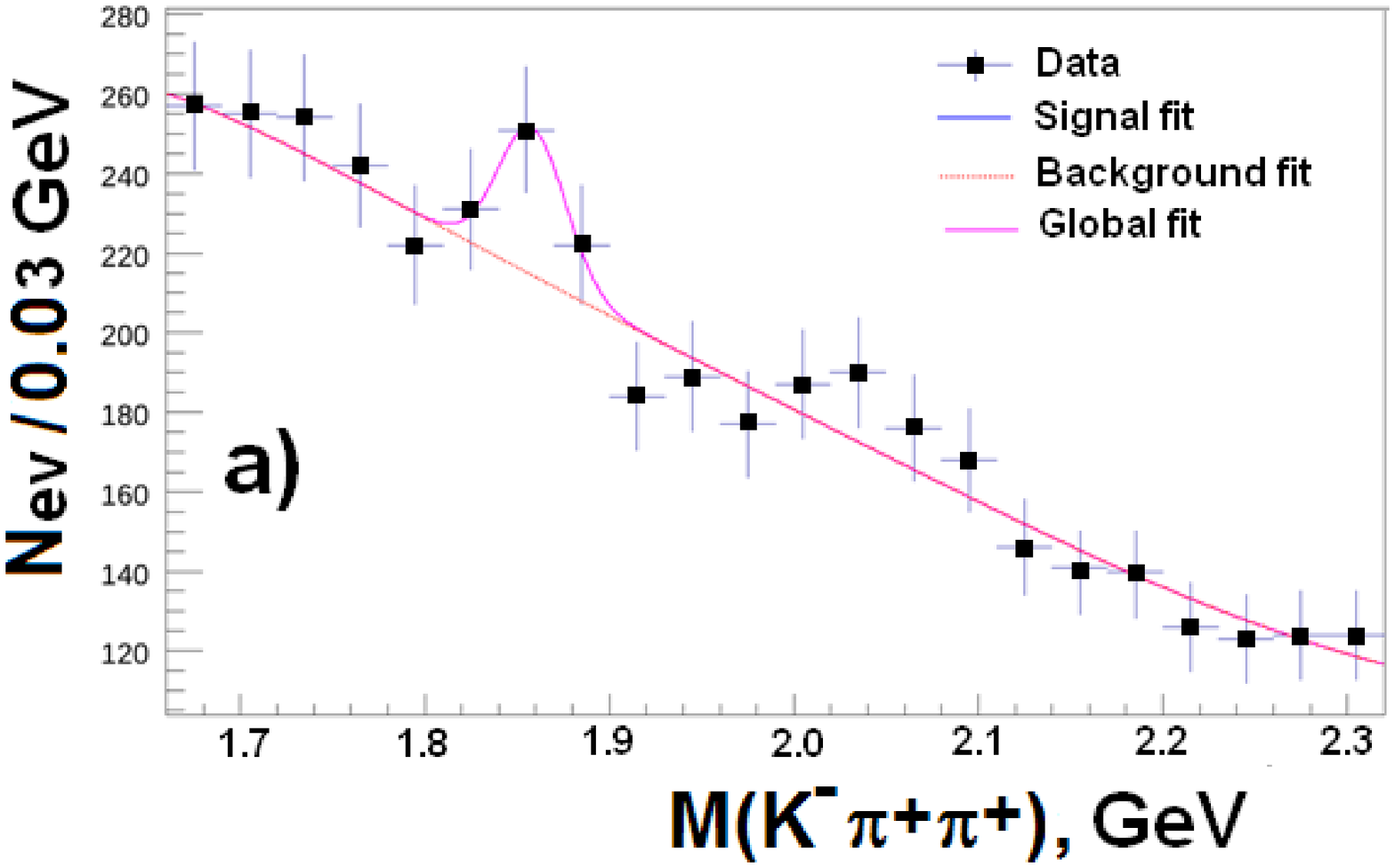}
\hspace{0.2in}
\includegraphics[scale=0.35]{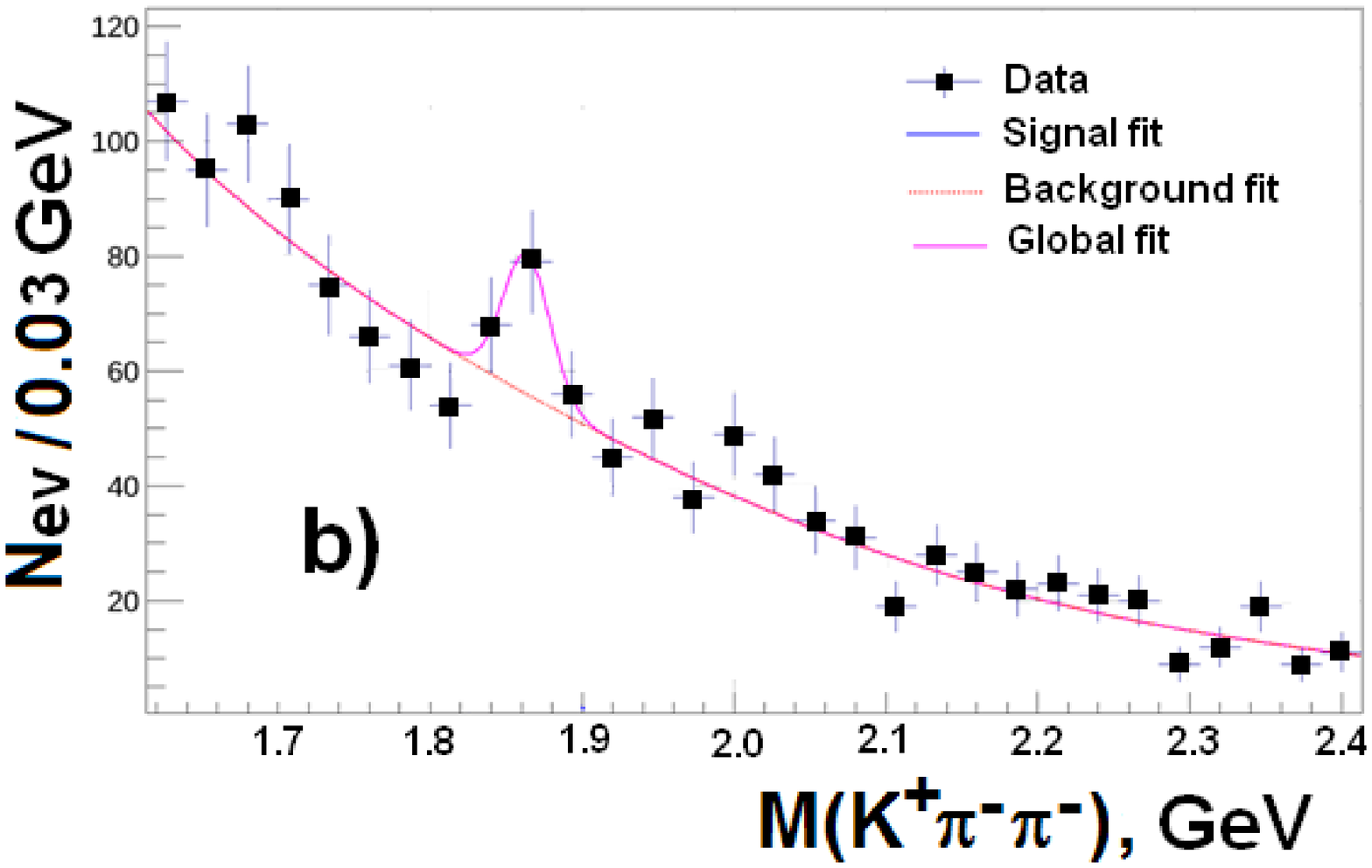}
\centering
\caption{Raw experimental effective-mass spectra of ($K^-\pi^+\pi^+$) (a) 
and ($K^+\pi^-\pi^-$) systems (b).}
\end{figure*}

\section{Modeling and optimization of selection criteria}
GEANT3.21 \cite{GEANT} program with the description of all SVD-2 components was used for modeling 
and optimization of selection criteria for $D^\pm$ meson events. FRITIOF7.02 \cite{PYTHIA}
program was 
used as the generator of pA-interactions. 
At first 
step the background under signals was simulated using 10 million Monte-Carlo (MC) events 
without charm. 3-prongs secondary vertices were found in some 
events because of detector noise and feature of procession algorithm. Distributions 
of some characteristics (decay length ($L$), momentum ($P$) and 
Feynman variable ($X_F$)) of 3-prongs systems for MC-events and experimental background 
in interval of $D$ meson masses from fig. 1 ($M=1.86\pm3*0.02$ (GeV)) were compared. 
The proper decay length $L$ was calculated from the observed $L_{lab}$ as $L=L_{lab}*M/P$.
All distributions really reproduce experimental background conditions. Momenta of 
3-prongs systems lie above 7 GeV.

A half of million MC-events with $D^+ \rightarrow K^-\pi^+\pi^+$ decay were used 
for optimization of selection criteria. At the first the Dalitz-plot for ($K^-\pi^+\pi^+$) 
system was analyzed in $m_1=m(K^-\pi^+_1)$ and 
$m_2=m(K^-\pi^+_2)$ coordinates. 
All MC-events are grouped within an ellipse (fig. 2a).
In fig. 2b mass plot for experimental events with MC-events ellipse is presented. 
For events in the ellipse, dependences of event densities on $\phi$ angle are shown 
in fig. 2c. From the analysis of fig. 2c the following selection criteria were 
taken: $\phi < 200^\circ$, or $\phi > 340^\circ$ and $R_{ell} <$ 1.  
\begin{figure*}[h]
\includegraphics[scale=0.5]{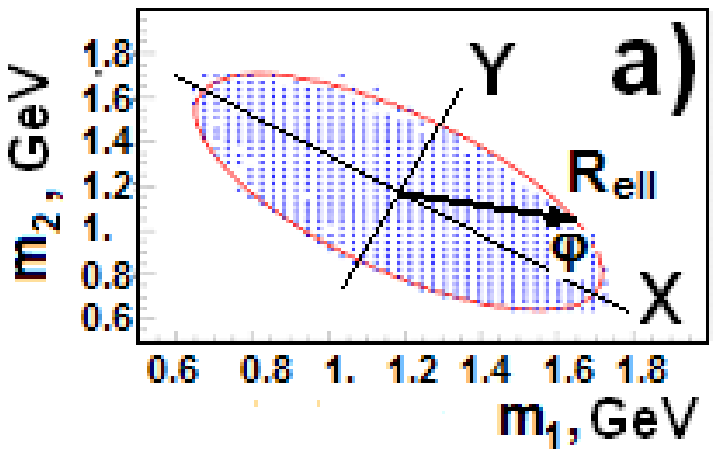}
\includegraphics[scale=0.35]{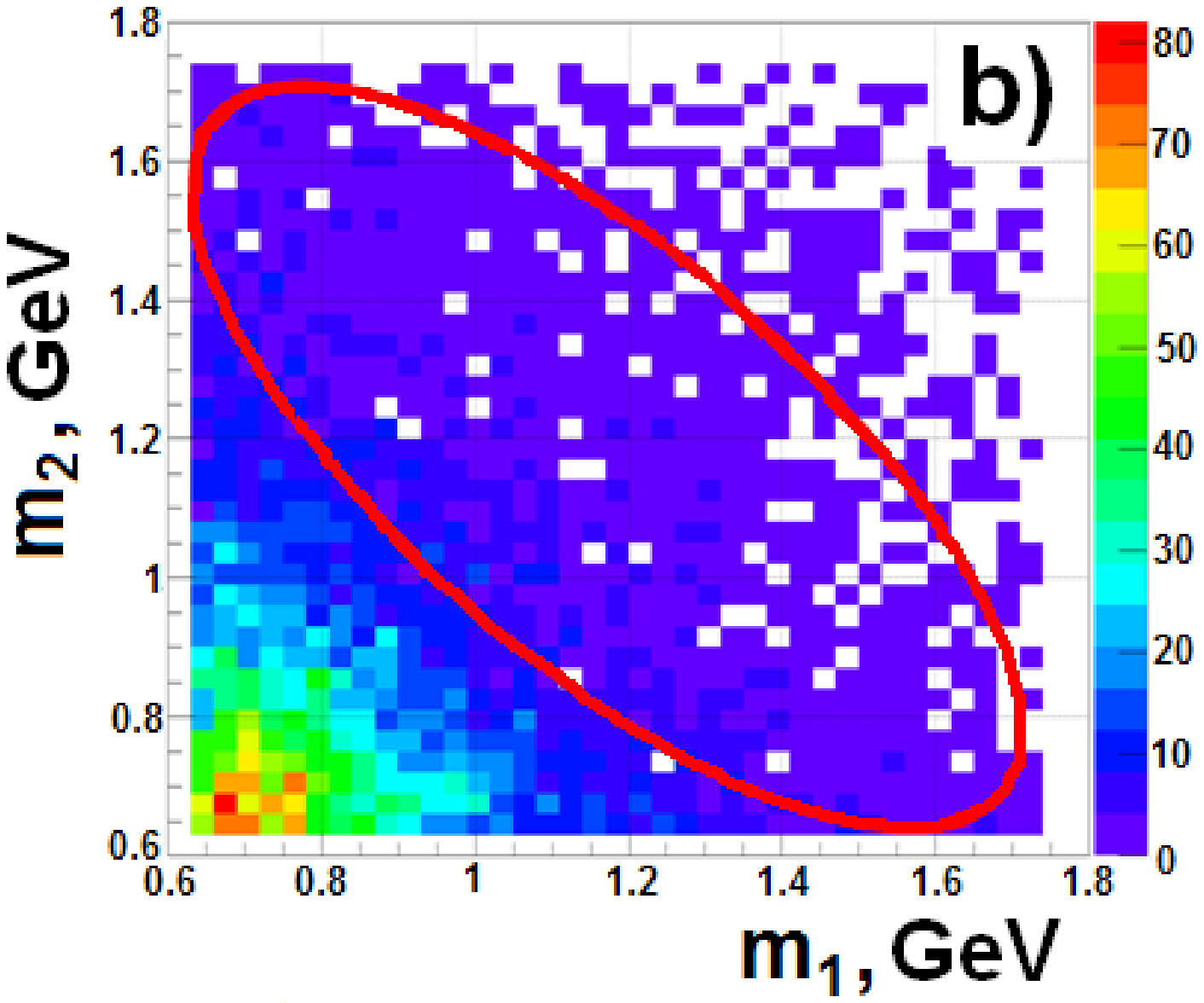}
\includegraphics[scale=0.3]{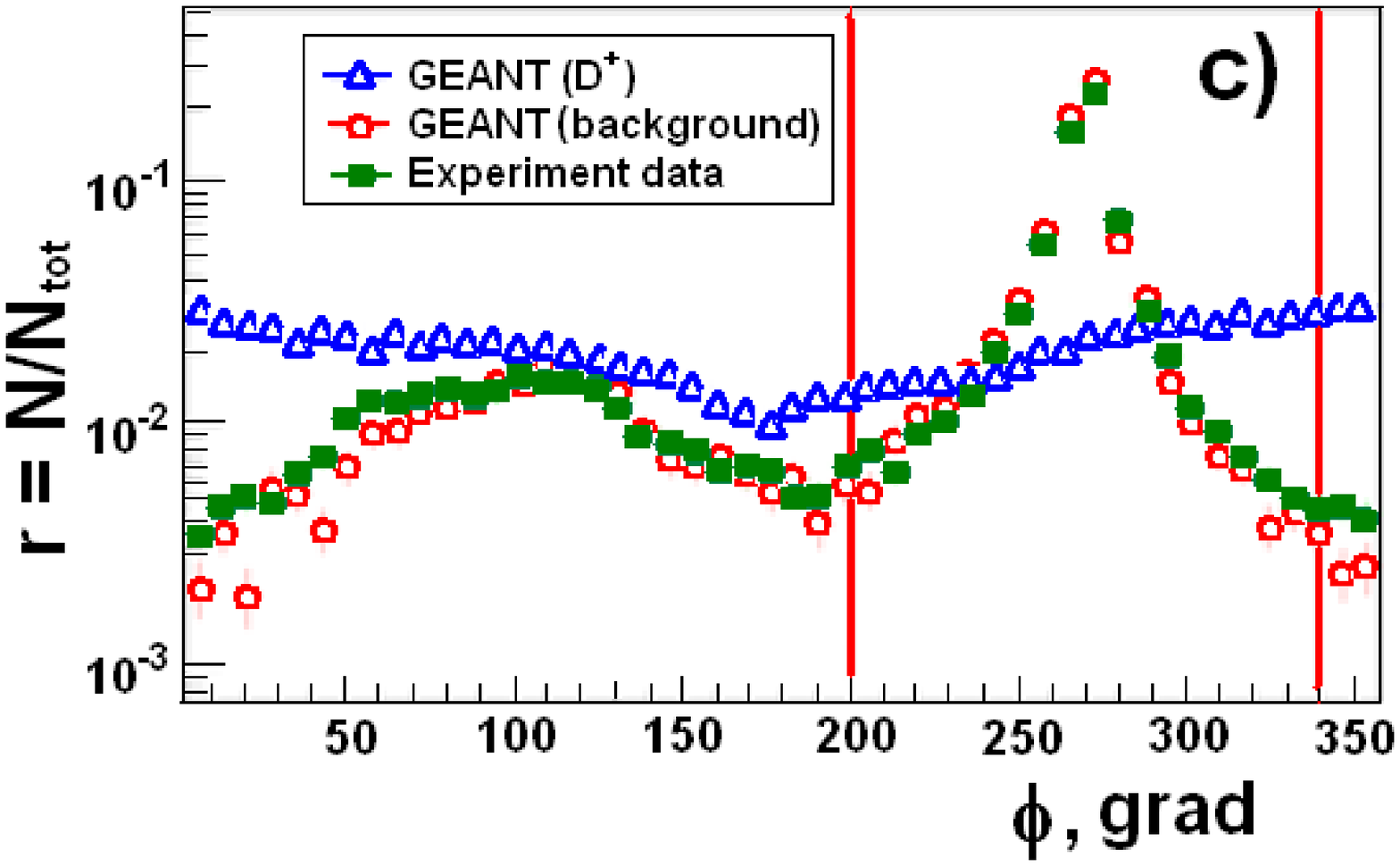}
\centering
\caption{a) Dalitz-plot for ($K^-\pi^+\pi^+$) system. b) MC-ellipse imposed on 
experimental Dalitz-plot.
c) Experimental and MC-events densities versus $\phi$.}
\end{figure*}

Another background occurs when charge track from primary vertex combined 
with $K^0$ decay vertex. 
To remove it another mass plot was considered:
in the same 3-prongs secondary vertex $K^-$ candidate was replaced with $\pi^-$ candidate.  
In fig. 3a the plot with two pions mass hypotheses is presented. 
The $K^0$ background lies in the lower part of the plot.
Events under the line
($M(\pi^+\pi^-)_{H1} + M(\pi^+\pi^-)_{H2} < C$)
have to be excluded. If $N_{cut}$ -- is the number of the rejected events and 
$N_{tot}$-- total number of events, then the share of rejected events 
$W=N_{cut}/N_{tot}$ depends on $C$. 
From the analysis 
of distributions in fig. 3b the cut parameter $C=$ 1.2 was taken for reduction of 
$K^0$ background. The $K^0$ background became practically unseen.
\begin{figure*}[h]
\includegraphics[scale=0.35]{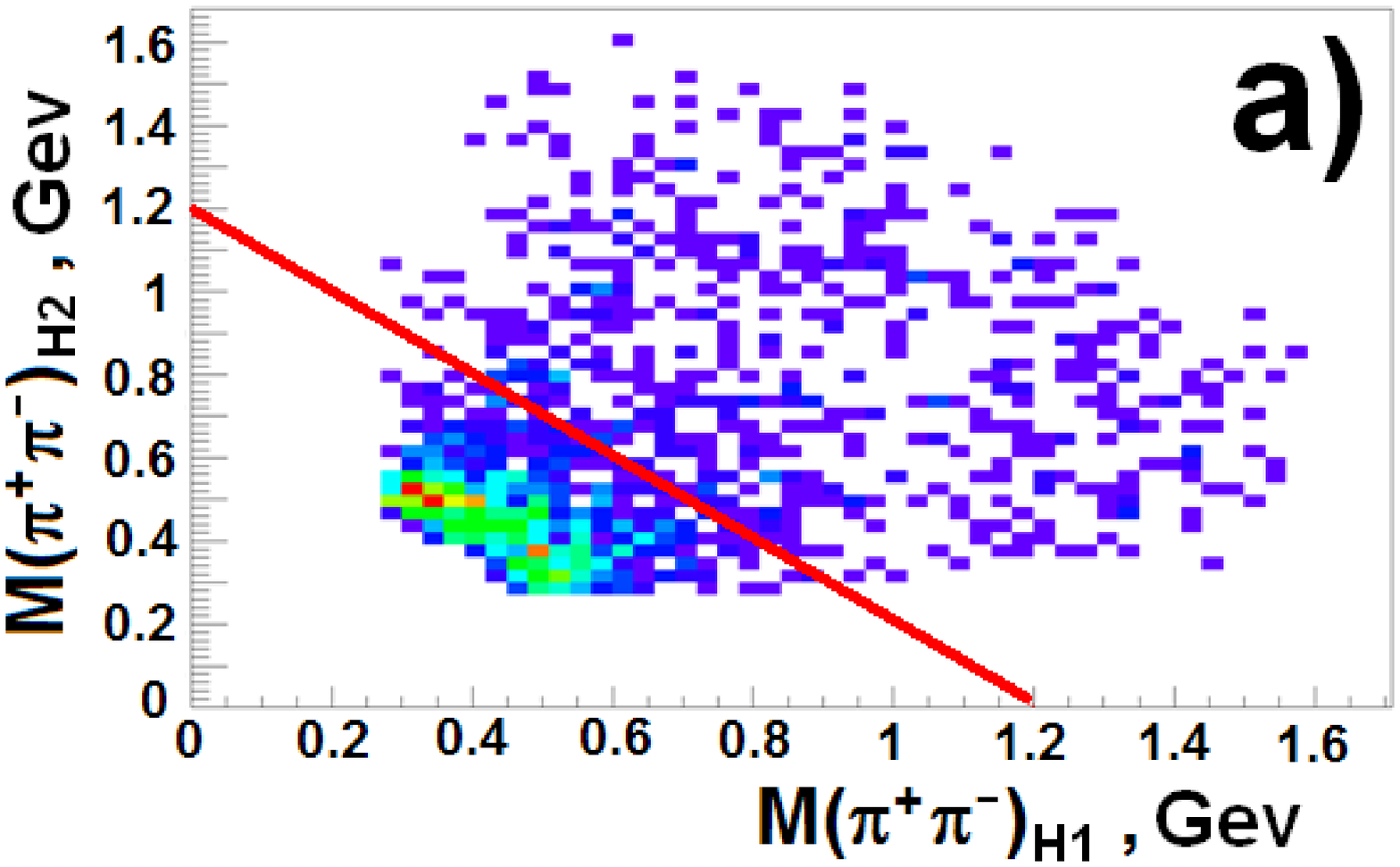}
\hspace{0.4in}
\includegraphics[scale=0.38]{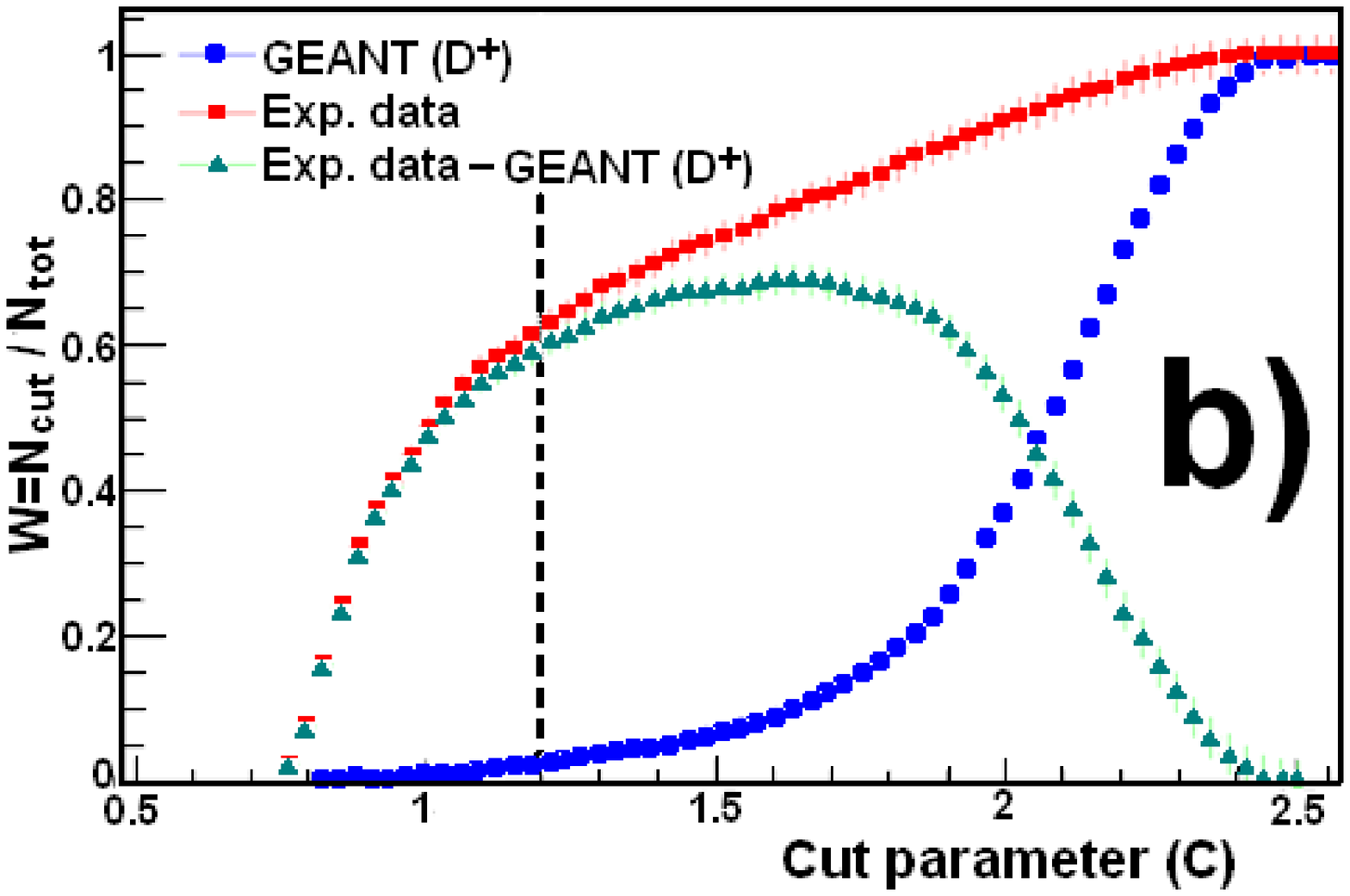}
\centering
\caption{a) The plot for ($\pi\pi\pi$) system.
b) Dependence of $W$ value on parameter $C$.}
\end{figure*}

Distributions for MC-events 
with $D^+$ and for experimental 
events with 3-prongs secondary vertex vs. proper decay length $L$ of 
($K^-\pi^+\pi^+$) system were got. From the analysis of these distributions the  
$L >$ 0.12 mm cut was introduced.

\section{The analysis of experimental events}
As a result of simulation the selection criteria for $D^+$ decays with the minimum 
background were taken:
1) $\phi(K^-\pi^+) < 200^\circ$, or $\phi(K^-\pi^+) > 340^\circ$ and $R_{ell} <$ 1;
2) $M(\pi^+\pi^-)_{H1} + M(\pi^+\pi^-)_{H2} <$ 1.2 GeV;
3) $L(K^-\pi^+\pi^+) >$ 0.12 mm.
The $D^+$ detection efficiency obtained after application of these criteria 
to 500000 MC-events is 1.4\verb|%|. The same criteria were applied to 
500000 MC-events with $D^-$ decays. Detection efficiency for $D^-$ is 0.8\verb|%|. 
In figs. 4a and 4b the experimental mass spectra of ($K^-\pi^+\pi^+$) and ($K^+\pi^-\pi^-$) 
systems are presented. Signals of $D$ mesons were fitted by the sum of Gaussian 
function and 6-order polynomial background. The parameters of the fits for $D^+$
were: $\chi^2/$NDF = 7.4 / 12, prob = 0.8; signal from $D^+$ = 15.4 events; 
background under the signal = 16.6 events; $D^+$ mass = 1873$\pm$5 MeV; 
standard deviation = 12 MeV.  For $D^-$: $\chi^2/$NDF = 2.7 / 11, 
prob = 0.99; signal from $D^-$ = 15.3 events; background under the signal = 8.7 events; 
$D^-$ mass = 1863$\pm$8 MeV; standard deviation = 22 MeV. 
The measured values of charged $D$ mesons masses are near to PDG value (1869.6 MeV) 
within the errors. In the mass interval of $D$ mesons a $K^0$ background was not found.
\begin{figure*}[h]
\includegraphics[scale=0.35]{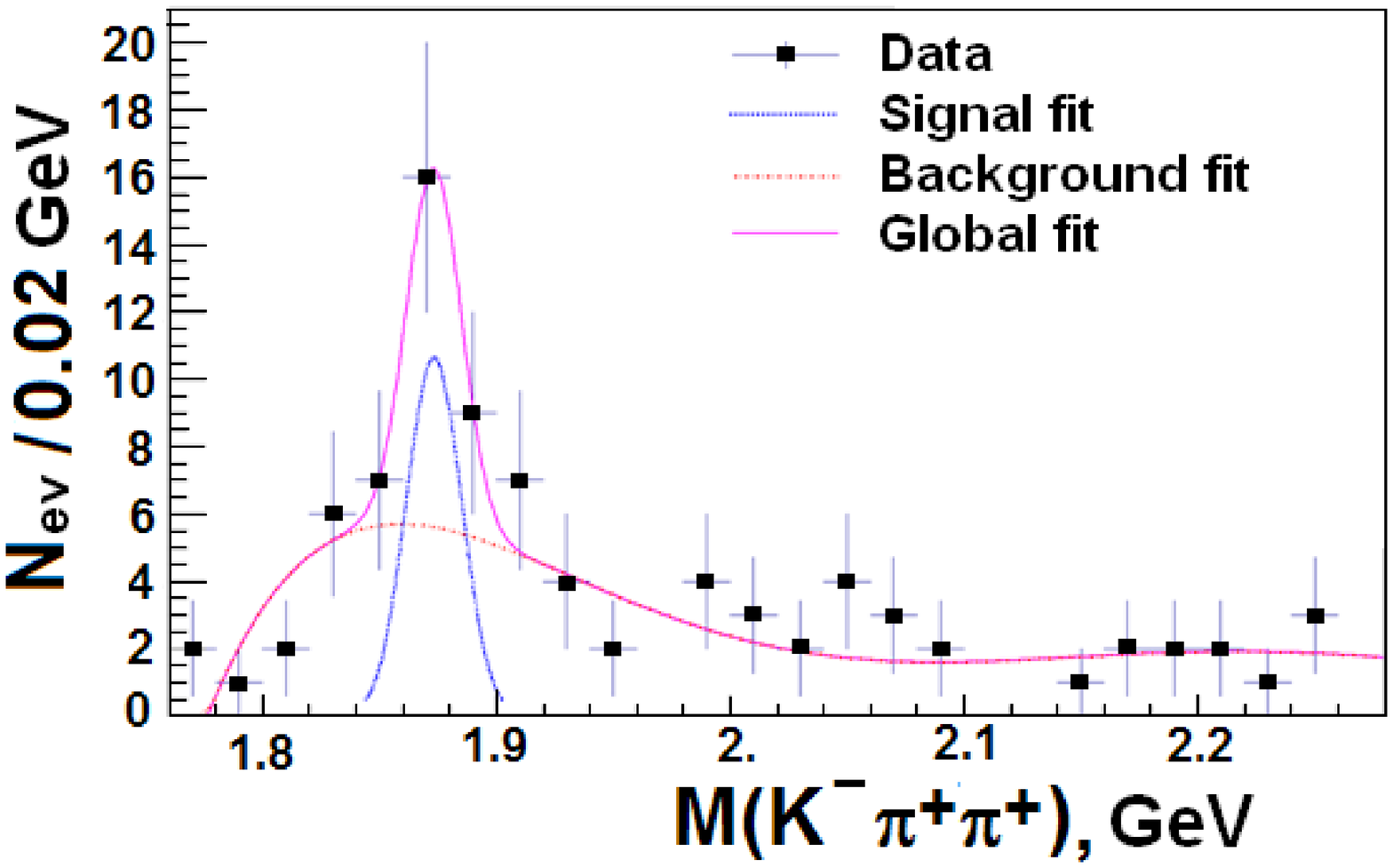}
\hspace{0.3in}
\includegraphics[scale=0.35]{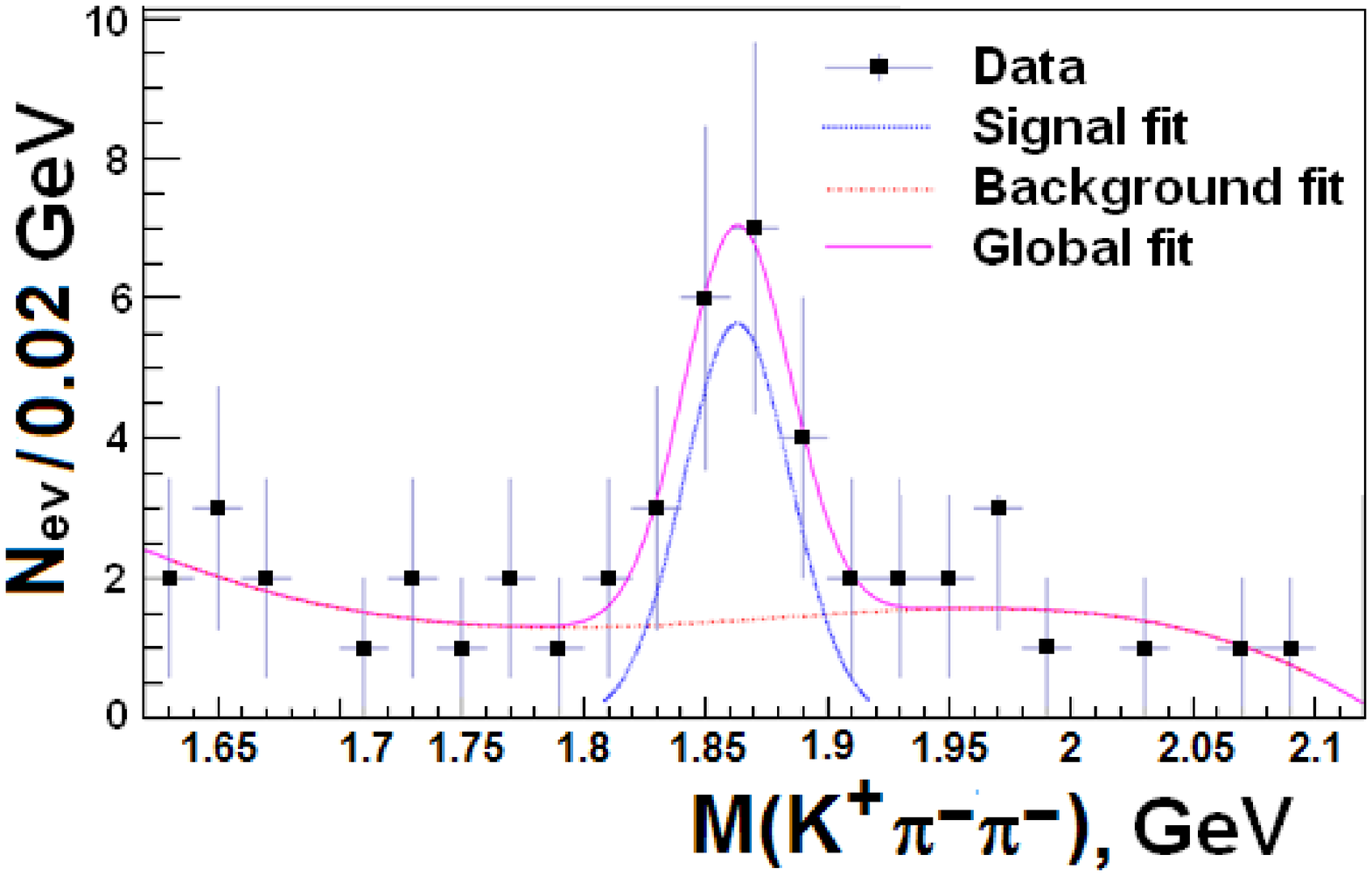}
\centering
\caption{Effective-mass spectra of ($K^-\pi^+\pi^+$) (a) and 
of ($K^+\pi^-\pi^-$) (b) systems.} 
\end{figure*}

Measuring $D$ mesons lifetime we can get another proof of observing charm particles. 
Events from signal slice ($M(D) \pm 2.5*\sigma$) were selected. 
The decay length distributions were constructed for them taking into account detection 
efficiency. The distributions were fitted by exponent. The background 
under a signal was estimated with help of distributions for MC-events. The measured 
values of the $c\tau$ parameters were 291 $\pm$ 75  $\mu$m for $D^+$ meson 
and 341 $\pm$ 88  $\mu$m for $D^-$ meson. They coincide with PDG value 
(311.8 $\mu$m) within the errors. Only statistical errors are here. 
The lifetime for events outside of signal areas considerably differ from these values.

\section{Inclusive cross sections and A-dependence}
The formula for calculation of cross sections was taken as:
\begin{center}
$N_s = [N^0(\sigma_DA^\alpha)/(\sigma_{pp}A^{0.7})]\times[($B$\varepsilon)/K_{tr}]$, where
\end{center}
$N_s$ -- number of events in a signal; 
$N_0$ -- number of events with pA-interactions in a target; 
$\sigma_D$ -- charm cross section; 
$A$ -- nuclear weight of target material (C, Si, Pb);
$\alpha$ -- parameter of A-dependence for charm cross section (= 0.7 for background); 
$\sigma_{pp}$ -- inelastic pp-interactions cross section at 70 GeV (= 31440 mb); 
B -- branching ratio of $D^\pm \rightarrow K\pi\pi$ decay (= 0.094);
$\varepsilon$ -- detection efficiency for $D$ mesons ($\varepsilon(D^+)$ = 0.014, 
$\varepsilon(D^-)$ = 0.008);
$K_{tr}$ = 0.57 (trigger efficiency \cite{PAN73} after specification);

With the expressions:
$C_D = [N^0/(\sigma_{pp}A^{0.7})]\times[($B$\varepsilon)/K_{tr}]$ and
$ln(N_s / C_D) = \alpha \times ln(A) + ln(\sigma_D)$
A-dependence of cross sections was received. The slope parameters of linear fits 
are: 1.02 $\pm$ 0.26 for $D^+$ and 1.04 $\pm$ 0.27 for $D^-$. 
The average values of inclusive cross sections (weighed on target materials) are: 
\begin{center}
$\sigma(D^+)$ = 1.2 $\pm$ 0.4(stat.) $\pm$ 0.2(syst.) ($\mu$b/nucleon),\\
$\sigma(D^-)$ = 1.9 $\pm$ 0.6(stat.) $\pm$ 0.4(syst.) ($\mu$b/nucleon).
\end{center}
The relative errors of the cross sections are: near 30\verb|%| from statistics 
and near 15\verb|%| from uncertainty of detection efficiency and of trigger factor 
calculations. 

\section{The ratios of charm meson yields}
In earlier paper \cite{PAN73} the estimation of open charm total cross section neutral $D$ mesons 
observations in pA-interactions at 70 GeV was obtained as:
\begin{center}
$\sigma(c\bar c)$ = 7.1 $\pm$ 2.4(stat.) $\pm$ 1.4(syst.) ($\mu$b/nucleon).\\
\end{center}
The cross sections of neutral charm mesons and anti-mesons were estimated as:
\begin{center}
$\sigma(D^0)$ = 2.5 $\pm$ 0.8(stat.) $\pm$ 0.5(syst.) ($\mu$b/nucleon),\\
$\sigma(\bar D^0)$ = 4.6 $\pm$ 1.6(stat.) $\pm$ 0.9(syst.) ($\mu$b/nucleon).
\end{center}
Table 1. Yields of D mesons and their ratios.
\begin{center}
\small{
\begin{tabular}{|l|c|c|c|c|c|c|c|} \hline
\multicolumn{1}{c}{Mesons}&\multicolumn{1}{c}{PYTHIA}&\multicolumn{3}{c}{FRITIOF}&
\multicolumn{1}{c}{SVD-2}&\multicolumn{2}{c}{Other experiments} \\
\multicolumn{1}{c}{}&\multicolumn{1}{c}{pp-int.}&\multicolumn{3}{c}{pA-interactions}&
\multicolumn{1}{c}{pA-int.}&\multicolumn{2}{c}{pA-interactions} \\ \hline
&&C&Si&Pb&&NA-27 \cite{PL189}&HERA-B \cite{EPJ52}  \\ \hline
$D^0$ & 0.28& 0.48&0.51&0.55&0.35$\pm$0.16&0.57$\pm$0.08&0.44$\pm$0.18 \\ \hline
$\bar D^0$ & 0.74& 0.60&0.59&0.58&0.65$\pm$0.31&0.43$\pm$0.09&0.54$\pm$0.23 \\ \hline
$D^+$ & 0.13& 0.28&0.29&0.29&0.16$\pm$0.07&0.31$\pm$0.06&0.19$\pm$0.08 \\ \hline
$D^-$ & 0.24& 0.28&0.27&0.28&0.27$\pm$0.17&0.34$\pm$0.06&0.25$\pm$0.11 \\ \hline
$D^0 / \bar D^0$&0.38&0.80&0.86&0.95&0.54$\pm$0.25&1.33$\pm$0.25&0.81$\pm$0.23 \\ \hline
$D^+ / D^-$ & 0.54& 1.0&1.1&1.0&0.59$\pm$0.20&0.92$\pm$0.21&0.76$\pm$0.22 \\ \hline
$D^\pm/$&0.36&0.51&0.51&0.5&0.44$\pm$0.24&0.65$\pm$0.21&0.46$\pm$0.18 \\
$(D^0+\bar D^0)$&&&&&&& \\ \hline
$D^+ / D^0$ & 0.18& 0.56&0.52&0.46&0.59$\pm$0.21&0.54$\pm$0.11&0.44$\pm$0.12 \\ \hline
$D^- / \bar D^-$ & 0.32& 0.47&0.46&0.48&0.42$\pm$0.26&0.78$\pm$0.19&0.47$\pm$0.14 \\ \hline
\end{tabular}
}
\end{center}

\begin{figure*}[h]
\includegraphics[scale=0.45]{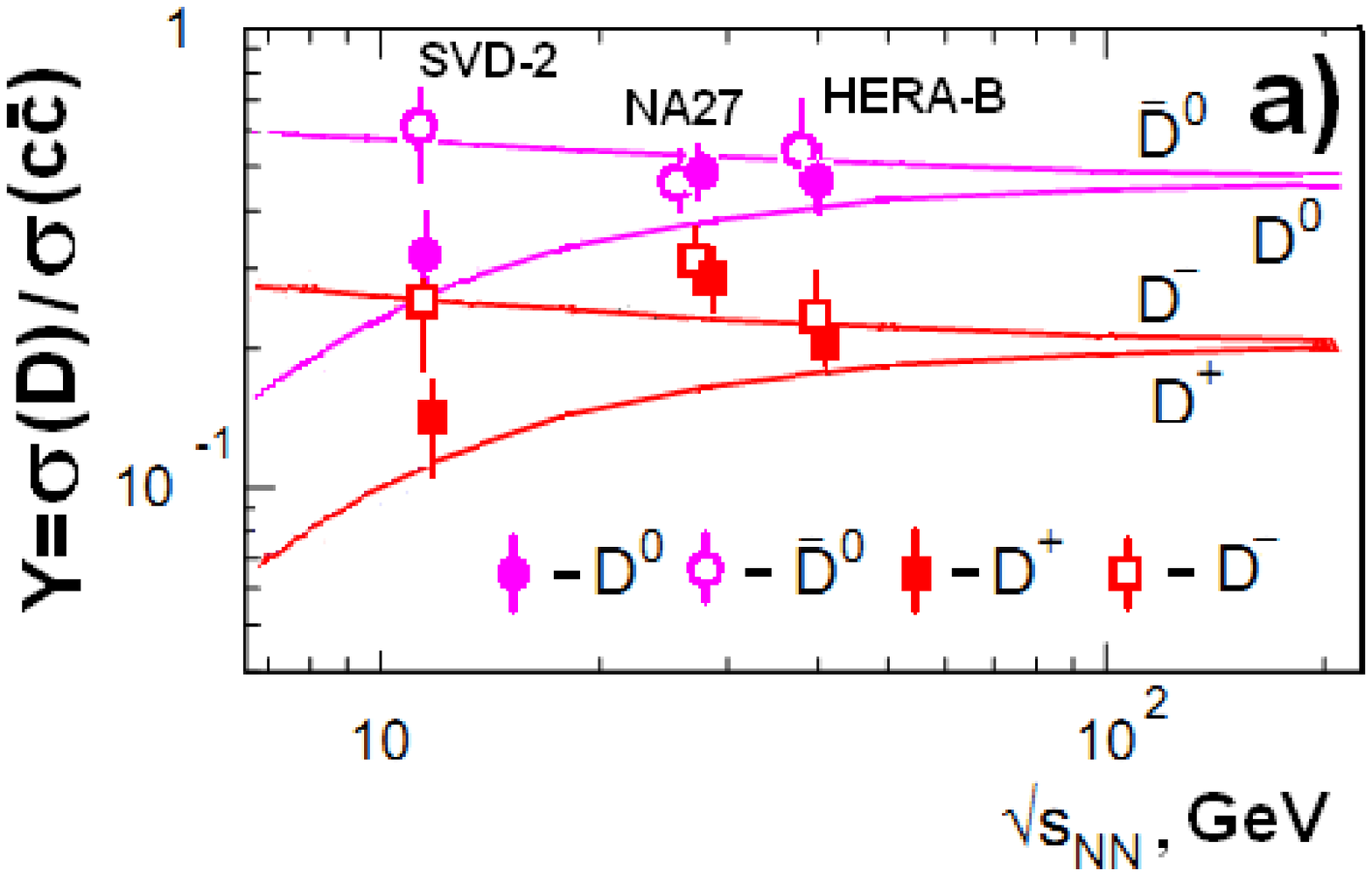}
\hspace{0.5in}
\includegraphics[scale=0.35]{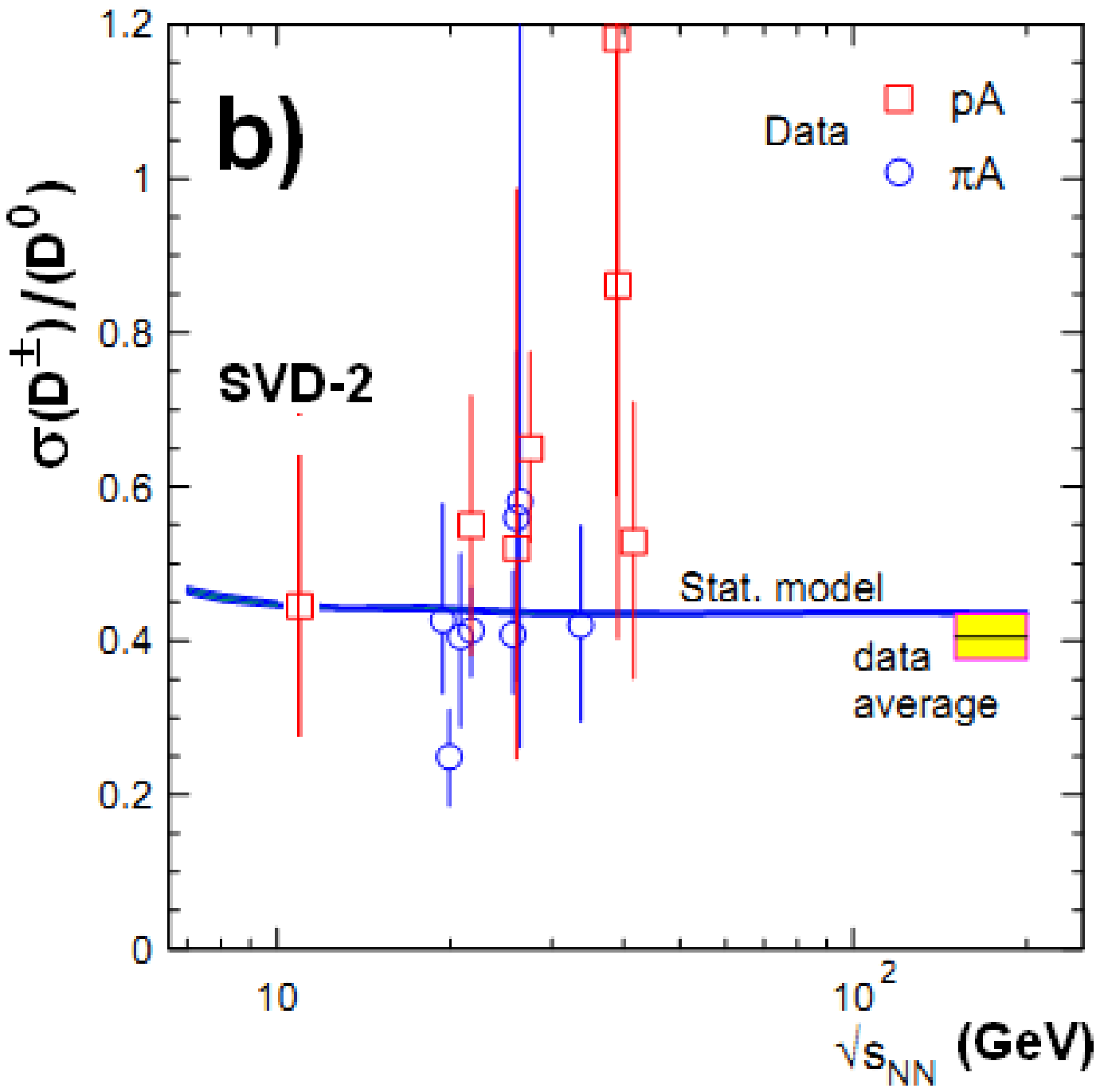}
\centering
\caption{a) Yields of D-mesons. b) Ratios of ($D^++D^-$) and ($D^0+\bar D^0$) cross 
sections. } 
\end{figure*}

In table 1 yields of $D$ mesons and their ratios using SVD-2 results and data 
of other experiments for pA-interactions are presented. The yields received from 
the PYTHIA and FRITIOF programs at our energy are also given. 
Fig. 5a shows that yields of mesons are decreasing with drop of energy, but 
yields of anti-mesons are increasing. The difference in the yields of particles 
and antiparticles was observed for the first time in a nA-interactions at average 
neutron beam energy 43 GeV in BIS-2 experiment \cite{ZP37}. 
In this experiment the decays of antiparticles ($\bar D^0$ and $D^-$) were detected, but 
the decays of particles ($D^0$ and $D^+$) were not found. Cross sections of particles 
production might appear below the sensitivity threshold in this experiment. 
In fig. 5b the ratios of cross sections of charged and neutral $D$ mesons from 
paper \cite{PR433} 
and the present result are shown. The results are compared to the predictions of 
statistical hadronization model \cite{Andronic}.

\section{Conclusion}
In SERP-E-184 experiment at SVD-2 setup (Protvino, Russia) $D^\pm$ mesons signals 
were obtained in effective-mass spectra of 3-prongs ($K\pi\pi$) systems in 
pA-interactions at 70 GeV. The selection criteria of events with open charm 
production were optimized using detailed simulation with FRITIOF7.02 and GEANT3.21 
programs. Inclusive cross sections of $D^\pm$ mesons production were estimated 
at near threshold energy.
The SVD-2 active target with plates of different materials (C, Si, Pb) allowed to measure 
the A-dependence parameters of cross sections for $D^\pm$ mesons production.
The yields of $D$ mesons and their ratios in comparison with data of other 
experiments and theoretical predictions were estimated. Experimental data showed 
the changes in $D$ mesons yields with a decrease of pA-interaction energy. These 
results are close to the predictions of a statistical hadronization model.

\end{document}